\documentclass[twocolumn,showpacs,preprintnumbers,prd,superscriptaddress,nofootinbib]{revtex4}
\usepackage{epsfig}
\usepackage{graphicx}
\usepackage{amsmath,amssymb,amsfonts}
\usepackage{array}
\usepackage{url}
\usepackage{hyperref}
\usepackage{lineno}
\usepackage{xspace}

\usepackage{graphics}
\usepackage{graphicx}
\begin{document}
\def\jour#1#2#3#4{{#1} {\bf#2}, #4 (19#3)} 
\def\jou2#1#2#3#4{{#1} {\bf#2}, #4 (20#3)} 
\def\PRL{{Phys. Rev. Lett. }}
\def\EPJA{{Eur. Phys. J.  A }}
\def\EPL{{Europhys. Lett.}}
\def\PRv{{Phys. Rev. }}
\def\PRC{{Phys. Rev.  C }}
\def\PRD{{Phys. Rev.  D }}
\def\JAP{{J. Appl. Phys. }}
\def\AJP{{Am. J. Phys. }}
\def\NIMA{{Nucl. Instr. and Meth. Phys. A }}
\def\NPA{{Nucl. Phys. A }}
\def\NPB{{Nucl. Phys.  B }}
\def\NPBP{{Nucl. Phys.  B, Proc. Suppl. }}
\def\NJP{{New J.  Phys. }}
\def\EPJC{{Eur. Phys. J.  C }}
\def\PLB{{Phys. Lett. B }}
\def\PHY{{Physics }}
\def\MPLA{{Mod. Phys. Lett. A }}
\def\PRp{{Phys. Rep. }}
\def\ZPC{{Z. Phys. C }}
\def\ZPA{{Z. Phys. A }}
\def\PPNP{{Prog. Part. Nucl. Phys. }}
\def\JPG{{J. Phys. G }}
\def\CPC{{Comput. Phys. Commun. }}
\def\APP{{Acta Physica Pol. B }}
\def\AIP{{AIP Conf. Proc. }}
\def\JHEP{{J. High Energy Phys. }}
\def\PSC{{Prog. Sci. Culture }}
\def\NCim{{Nuovo Cim. }}
\def\SNC{{Suppl. Nuovo Cim. }}
\def\SJNP{{Sov. J. Nucl. Phys. }}
\def\SPJ{{Sov. Phys. JETP }}
\def\JLet{{JETP Lett.}}
\def\PTP{{Prog. Theor. Phys. }}
\def\PTPS{{Prog. Theor. Phys. Suppl. }}
\def\IANSF{{Izv. Akad. Nauk SSSR: Ser. Fiz. }}
\def\JPCS{{J. Phys. Conf. Ser. }}
\def\AHEP{{Adv. High Energy Phys. }}
\def\IJMPE{{Int.  J. Mod. Phys. E }}
\def\ARNPS{{Ann. Rev. Nucl. Part. Sci. }}

\def\ct{\cite}
\def\sNN{\sqrt{s_{N\!N}}}
\def\sNNq{s_{N\!N}}
\def\sppq{s_{pp}}
\def\spp{\sqrt{s_{pp}}}
\def\eNN{\varepsilon_{N\!N}}
\def\pbp{{\bar p}p}

\def\col{Collab.}
\def\bi{\bibitem}
\def\ea{{\sl et al.}}
\def\eg{{\sl e.g.}}
\def\vrs{{\sl vs.}}
\def\ie{{\sl i.e.}}
\def\va{{\sl via}}
\def\nopar{\noindent}
\def\bi{\bibitem} 
\def\lssim{\stackrel{<}{_\sim}}
\def\gtsim{\stackrel{>}{_\sim}}
%


\title{
 Centrality  dependence of midrapidity density   
from GeV to TeV heavy-ion collisions\\  in the 
  effective-energy  universality picture of hadroproduction
}
\author{Edward K. G. Sarkisyan}
\email {sedward@cern.ch}
\affiliation{Experimental Physics Department, CERN, 1211 Geneva 23, 
Switzerland}
\affiliation{Department of Physics, The University of Texas at
Arlington, Arlington, TX 76019, USA}
\author{Aditya Nath Mishra}
\email {Aditya.Nath.Mishra@cern.ch}
\affiliation{Discipline of Physics, School of Basic Science, Indian
Institute of Technology, Indore 452020, India}
\author{Raghunath Sahoo}
\email{Raghunath.Sahoo@cern.ch}
\affiliation{Discipline of Physics, School of Basic Science, Indian
Institute of Technology, Indore 452020, India}
\author{Alexander S. Sakharov}
\email{Alexandre.Sakharov@cern.ch}
\affiliation{Experimental Physics Department, CERN, 1211 Geneva 23, 
Switzerland}
\affiliation{Department of Physics, New York University, New York, NY
10003, USA}
\affiliation{Physics Department, Manhattan College, Riverdale, NY 10471,
USA}

 \begin{abstract} 
 The dependence on centrality, or on the number of nucleon participants, 
of the midrapidity density of charged particles measured in heavy-ion 
collisions at the 
collision energy of about 20 
GeV  at 
RHIC to the highest LHC energy of 5 TeV is 
 investgated
 within the recently proposed effective-energy approach. 
 This approach  
 relates
multihadron production in different types of collisions
 by combining,
 under the proper scaling of the collision energy,
 the constituent quark
 picture
 with Landau relativistic hydrodynamics.
  The  measurements are shown to be well described 
  based on  the similarity of multihadron production 
  process in 
 (anti)proton-proton interactions and heavy-ion collisions
driven by the centrality-dependent effective energy of participants.

 \end{abstract}

\maketitle

 Recently, we have shown that
the multiplicity \ct{edwardPRD}
and  midrapidity density
\ct{edward2} data  from 
heavy-ion collisons in the collision energy range of  several orders of 
magnitude are well described in the framework of the picture of the 
dissipating 
effective energy of constituent quark participants \ct{edwarda,edward}, 
or, for brevity, 
the effective-energy approach.
 In this paper, we show that the recent measurements 
of the
centrality dependence of the pseudorapidity midrapidity density of charged 
particles in PbPb collisons 
by ALICE 
 \ct{alice5020},
at the highest center-of-mass (c.m.) energy 
 per 
nucleon, $\sNN$,
 ever reached, namely at $\sNN=5.02$~TeV,  
are also well described using the effective-energy 
approach.

 This approach interrelates the particle 
production process in different 
types of 
collisions \ct{edwarda}, as briefly described below.
 Within 
 such a 
 picture
the process of particle production is quantified in terms of the 
amount of {\it effective} energy
deposited by interacting constituent quark participants into the small
Lorentz-contracted volume formed at the early stage of a collision.
The approach 
 considers
the Landau
 relativistic
hydrodynamic approach
 to   
 multiparticle production 
 \ct{Landau}
 employed in the framework of constituent (or dressed) quarks, in 
 accordance with the additive quark model \ct{additq,constitq}.
 Then, in $pp/\pbp$ collisions, a single constituent quark from  
 each nucleon is assumed
  to   contribute
 in a collision.
 The remaining quarks
 are treated as spectators
 resulting
 in 
 formation of leading particles 
 carrying away a  significant part of the collision energy.
 On the contrary, in
 the head-on heavy-ion collisions, 
 all three constituent quarks from each of the participating nucleons 
 are considered to contribute
 so that the whole energy of the 
 nucleons 
 becomes available for the 
 particle production.  
 Thus, 
 the 
 bulk 
 measurements
 in 
 head-on heavy-ion collisions 
 at 
 $\sqrt{s_{NN}}$
 are 
  treated to be similar to 
  those from
 $pp/\pbp$ collisions 
 at the 
 properly
 scaled  c.m. energy $\sqrt{s_{pp}}$, 
 \ie\ 
 at 
 $\sqrt{s_{pp}} \simeq 
 3\sqrt{s_{NN}}$.

 Being put together, the above-discussed ingredients
 lead to 
 the relationship between charged particle 
 (pseudo)rapidity 
 density 
per participant pair at midrapidity,
 $\rho(\eta)=(2/N_{\rm{part}})dN_{\rm{ch}}/d\eta$ 
 ($\it{\eta} \approx $~0), in heavy-ion 
 collisions  and in  $pp/\pbp$  
 interactions:

 \begin{equation}
 \frac{\rho(0)}{\rho_{pp}(0)} = 
 \frac{2N_{\rm{ch}}}{N_{\rm{part}}\, N_{\rm{ch}}^{pp}} 
 \, \sqrt{\frac{L_{pp}}{L_{N\!N}}} \, ,\:\:\:\:\:
 \spp=3\sNN\, .
 \label{eqn1}
 \end{equation}
 Here, 
 $N_{\rm{ch}}$ and 
 $N_{\rm{ch}}^{pp}$ are the (total) mean multiplicities in nucleus-nucleus 
and nucleon-nucleon collisions, respectively, and $N_{\rm {part}}$ is the 
number of 
 nucleon
 participants. The relation of the pseudorapidity density and the 
mean multiplicity is applied in its Gaussian form as obtained in Landau 
hydrodynamics. The factor $L$ is defined as $L = {\ln}({\sqrt{s}}/{2m})$. 
According to the approach considered, $m$ is the proton mass, $m_{p}$, in 
nucleus-nucleus collisions and
 the constituent quark mass in $pp/\pbp$ collisions 
 set
 to  $\frac{1}{3}m_{{p}}$.
  Such an universality was found to correctly predict \ct{edward} the 
value of the midrapidity density in $pp$ interactions measured at LHC TeV 
energies \ct{CMSwe}.


%

 In the 
 further development  \ct{edward2}, 
 one 
 considers 
 the obtained relation, Eq. (\ref{eqn1}), 
 in
terms of centrality.
 The centrality is 
 regarded as the degree  of the overlap of the  volumes of the 
 two colliding  nuclei, 
 characterized  by the impact parameter,
 and
is closely 
related to the number of nucleon participants.
  Hence,
 the centrality  is  
 related to the amount of the energy 
 released in the collisions, \ie\  to the effective energy, $\eNN$.
 In the framework of the proposed approach, the effective energy 
 can be 
defined as 
a fraction of the c.m. energy available in a collision according to the 
centrality, $\alpha$:

\begin{equation}
\eNN = \sqrt{s_{NN}}(1 - \alpha).
\label{Eeff}
\end{equation}

Then, 
after taking into account the energy scaling,
Eq.~(\ref{eqn1})  
 reads
for the effective c.m.  energy 
$\eNN$:
 
\begin{eqnarray}
 \nonumber
&\rho(0)& = \rho_{pp}(0)\,
\frac{2N_{\rm{ch}}}{N_{\rm{part}}\, N_{\rm{ch}}^{pp}}
\, \sqrt{1 -
\frac{2 \ln 3}{\ln (2m_p/\eNN)}}\,,  \\
& & \eNN=\spp/3\,,
\label{pmultc}
\end{eqnarray} 
 where
$N_{\rm{ch}}$
is the  multiplicity
 in central nucleus-nucleus collisions measured 
at  $\sNN=\eNN$, 
and 
 $\rho_{pp}(0)$ 
 and 
 $N_{\rm{ch}}^{pp}$ are taken at  $\spp=3\, \eNN$.


\begin{figure*}
 \begin{center}
\resizebox{0.7\textwidth}{!}{%
  \includegraphics{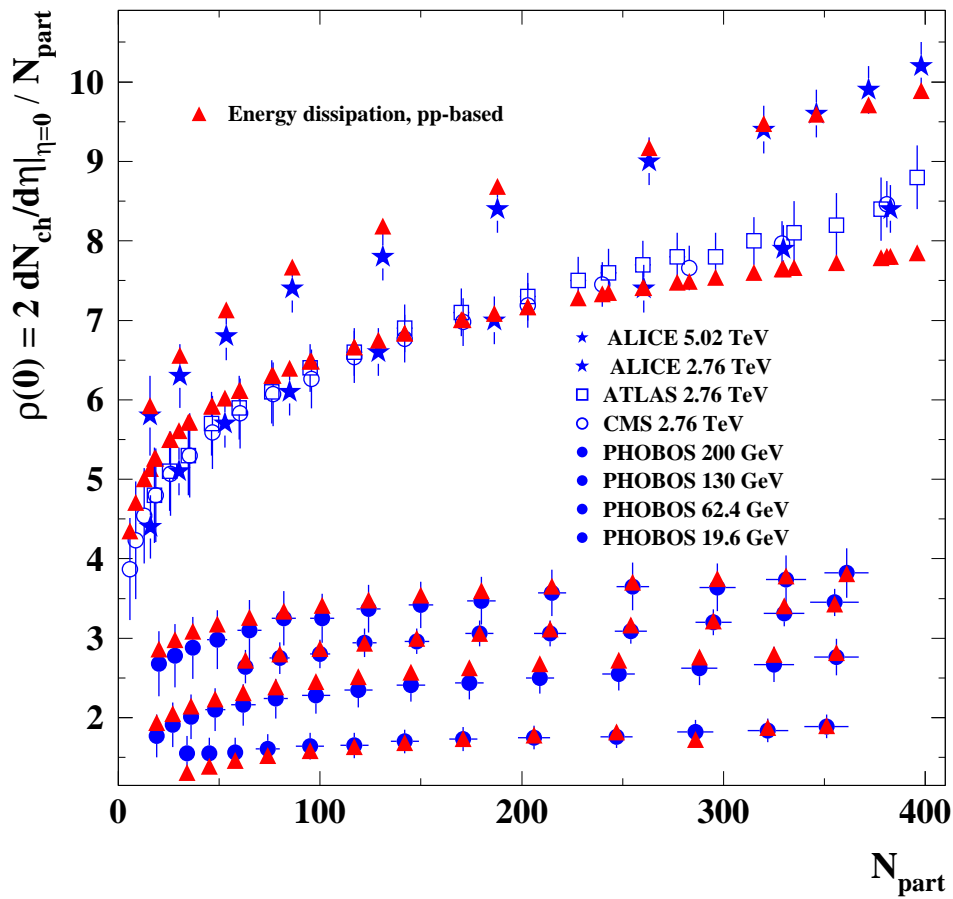}}
 \end{center}
\caption{
 The
charged particle pseudorapidity density
at midrapidity
per participant
pair as a function of the number of participants, $N_{\rm part}$.
 The
 solid
 circles show the 
 measurements from 
AuAu collisions at RHIC by
PHOBOS at $\sNN=19.6$ to 200~GeV
\ct{phobos-all} (bottom to top). The
 LHC measurements are
 from 
 PbPb collisions 
 by 
 ALICE  at $\sNN=2.76$~TeV  \ct{alice276c} 
 and 
 $\sNN=5.02$~TeV \ct{alice5020} (solid stars), 
 and by ATLAS 
\ct{atlas276c} (open squares) 
 and CMS 
\ct{cms276c}
(open circles) 
 at $\sNN=2.76$~TeV.
 The
 solid
 triangles show the calculations by
Eq.~(\ref{pmultc}) using $pp/\pbp$  data.
 }
\label{Fig1}       
\end{figure*}
%

 In Fig.~\ref{Fig1}, the $N_{\rm part}$-dependence of the charged particle 
pseudorapidity
 density $\rho(0)$, 
 measured in heavy-ion collisions at $\sNN$ from GeV c.m. energies by 
the PHOBOS experiment at RHIC \ct{phobos-all} to a 
 few TeV c.m. energies by the ALICE 
 \ct{alice5020,alice276c}, ATLAS \ct{atlas276c}, and CMS \ct{cms276c}
 experiments at LHC
 is compared with the calculations of Eq.~(\ref{pmultc}). 
  According to the
 consideration, the calculations are made at $\spp=3\, \eNN$.
 The midrapidity density $\rho_{pp}(0)$ and the multiplicity 
$N_{\rm{ch}}^{pp}$ are taken from the existing $pp/\pbp$ data 
\ct{pprev,pdg14}, and the $N_{\rm ch}$ values are taken from the heavy-ion 
collision data \ct{edwardPRD,alice-mult} where available, while
 otherwise 
the  corresponding experimental c.m. energy 
 fits
are applied.
 The calculations use the 
power-law $s_{pp}$-fits for $N_{\rm{ch}}^{pp}$ \ct{pprev} and for  
$\rho_{pp}$ at $\sqrt{s_{pp}}>$~53~GeV \ct{cms276c}, the linear-log fit 
\ct{pprev} for $\rho_{pp}$ at  $\sqrt{s_{pp}}\leq$~53~GeV, and the ``hybrid'' 
 $\sNNq$-fit \ct{edward2} for $N_{\rm ch}$.
 
 One can see that within this approach
 where the collisions are derived by the centrality-defined effective c.m. 
 energy $\eNN$, the 
 calculations are in very good overall agreement with the measurements 
independent of the collision energy.
   Similar results are obtained as the $N_{\rm part}$-dependence of the 
PHENIX \ct{phenix-all}, STAR \ct{star-all}, or CuCu PHOBOS \ct{phobos-all} 
measurements
 from RHIC 
 are used (not shown). Some slightly lower values 
 seen in the
 predictions
 compared to the data
for some low-$N_{\rm part}$, \ie\ for the most peripheral collisions,
at $\sNN=19.6$~GeV
 looks to be
due to the experimental limitations and the
extrapolation used in the reconstruction  for
the measurements in this region of very low multiplicity \ct{phobos-all}.
 This may  also  
 explain the $N_{\rm part}$-scaling of the data at $\sNN=19.6$~GeV in the 
 most peripheral region which does not
 follow the common trend of decreasing  
 observed
 at
 higher energies. 
 %
 The calculations obtained to be lower than the data
 for a few most central collisions at the LHC energy at $\sNN=2.76$~TeV 
and some slight deviations seen for the 5.02 TeV predictions can be 
 explained
 by yet to come measurements of $N_{\rm ch}^{pp}$ at energies above 
the top Tevatron energy of  $\spp=1.8$~TeV.
 
Recently, we have shown that, within the effective-energy approach, one 
describes as well the mean multiplicity data from heavy-ion collisions up 
to $\sNN=$2.76 TeV \ct{edwardPRD}. Moreover, the pseudorapidity 
distribution in the {\it full}-$\eta$ range, and not only the midrapidity 
density, are shown to be reproduced. The findings and a new 
energy-balanced limiting fragmentation scaling introduced in 
\ct{edwardPRD} elucidate the differences observed in the multiplicity and 
the midrapidity density centrality dependence as measured at RHIC and LHC.
 The description of the observed dependences suggest 
a possible change of the multihadron production mechanism in heavy-ion 
collisions when one moves to TeV energy heavy-ion collisions, where
the 
collisions seem to obey a head-on collision regime for all centralities.
 The midrapidity density is expected to increase with the number of 
 participants both at RHIC and 
 LHC as soon as the central-$\eta$ region 
is formed by considered to be centrally-colliding participants at the
c.m. energy of $\eNN$. This increase is shown by the 
measurements, see 
Fig.~\ref{Fig1}, and 
 is well described 
 by
 the approach 
discussed here. Similar to the midrapidity density, 
the multiplicity 
is also expected to demonstrate the increase with the number of 
participants. Such a  behaviour, observed at the TeV LHC energies, 
are shown to be well 
described by the effective-energy approach, and then 
treated to indicate the central collision regime independent of 
centrality.
  However, at RHIC, the multiplicity measurements show a constancy with 
the centrality, in contrast to the midrapidity behaviour at the same 
energies. This effect is shown to be due to the fact that, at RHIC 
energies, the multiplicity gets an additional contribution because of the 
difference between the collision energy and the effective energy shared by 
the interacting participants. This contribution is taken into account by 
the proposed energy-balanced limiting fragmentation within the 
effective-energy approach by considering the limiting fragmentation 
scaling in terms of the effective energy $\eNN$. This allows to well 
describe the multiplicity  and 
the pseudorapidity distribution for all energies 
 independent of centrality
 \ct{edwardPRD}.
 Additionally, 
 in \ct{edward2} the transverse-energy midrapidity densities are shown to 
demonstrate the complementarity of the head-on data and the centrality 
data in terms of the effectve energy, similar to that obtained  
for the midrapidity densities \ct{edward2} and the mean multiplicities 
\ct{edwardPRD}. 
 
 Interestingly, 
 this picture is shown as well  
  to successfully explain \ct{edwarda,edward}
 the similarity of the measurements in other collisions, such the 
 scaling between the charged particle mean multiplicity in $e^+e^-$ and 
$pp/\pbp$ collisions \ct{lep} and the universality of both the 
multiplicity and the midrapidity density measured in the most central 
nuclear collisions and in $e^+e^-$ annihilation \ct{phobos-sim}; see 
\ct{book} for discussion.
 In the latter case, colliding leptons are considered to be structureless 
and deposit their total energy into the Lorentz-contracted volume, 
similarly to nucleons in head-on nuclear collisions \ct{edward}. This is
 shown to be supported 
 by 
the observation made in  \ct{pprev} where 
 the multiplicity measurements in $pp/\pbp$ interactions up to TeV 
 energies are
  shown to be well reproduced by $e^+e^-$ data as soon as the inelasticity 
is set to $\approx$0.35, \ie\ effectively
  1/3 of the  hadronic interaction energy.
  For recent discussion on the universality of hadroproduction up to LHC 
energies, see \ct{pdg14};
 see also \ct{book,morepapers}.

 To summarize,
 the effective-energy dissipation approach based on the picture which 
combines the constituent quark model together with Landau relativistic 
hydrodynamics in view of the universality of the multihadron production in 
hadronic and nuclear collisions is shown to well describe the data from 
heavy-ion collisions within the c.m. energy of several orders of 
magnitude; in particular the centrality dependence of the midrapidity 
density of charged particles measured in heavy-ion collisions up to 5.02 
TeV is shown to be well reproduced. Future measurements at higher 
energies in different types of collisions are  crucial in order to 
 clarify 
 the underlying features of hadroproduction mechanism.
 \\

 The work of Alexander Sakharov is partially supported by the US National
 Science Foundation under Grants No.PHY-1205376 and No.PHY-1402964.

\end{document}